%% file: main.tex
\documentclass[sigconf,natbib=true,screen=true]{acmart}

\renewcommand\footnotetextcopyrightpermission[1]{} 
\input{packages}
\input{definitions}
\input{meta}

\begin{document}
\input{authors}

\title{PRADA: Practical Black-Box Adversarial Attacks against Neural Ranking Models}

\begin{abstract}
\blfootnote{Preprint. Under review.}
Neural ranking models (NRMs) have shown remarkable success in recent years, especially with pre-trained language models. However, deep neural models are notorious for their vulnerability to adversarial examples. Adversarial attacks may become a new type of web spamming technique given our increased reliance on neural information retrieval models. Therefore, it is important to study potential adversarial attacks to identify vulnerabilities of NRMs before they are deployed.

In this paper, we introduce the Word Substitution Ranking Attack (WSRA) task against NRMs, which aims to promote a target document in rankings by adding adversarial perturbations to its text. 
We focus on the decision-based black-box attack setting, 
where the attackers cannot directly get access to the model information, 
but can only query the target model to obtain the rank positions of the partial retrieved list.  
This attack setting is realistic in real-world search engines. 
We propose a novel Pseudo Relevance-based ADversarial ranking Attack method (PRADA) that learns a surrogate model based on Pseudo Relevance Feedback (PRF) to generate gradients for finding the adversarial perturbations. 

Experiments on two web search benchmark datasets show that PRADA can outperform existing attack strategies and successfully fool the NRM with small indiscernible perturbations of text. 
\end{abstract}



\maketitle

\section{introduction}

Ranking models are central to information retrieval (IR) research.
With the advance of deep neural networks, 
we are witnessing a rapid growth in neural ranking models (NRMs) \cite{drmm,Duet,dai2019deeper,onal-neural-2018}, achieving new state-of-the-art results in learning query-document relevance patterns. 
Recent research has explored pre-trained language models (e.g., BERT \cite{BERT} and ELMo \cite{elmo}) in the context of document ranking, and shown that they can achieve remarkable success on a variety of search tasks \cite{ma2021b,joshi2020spanbert,gu2020speaker}. 
The impact of pre-trained models is not limited to academic research. 
In industry, BERT and, more generally, transformers are being put to practical usage \citep[see, e.g.,][]{lin-2021-pretrained}. 

\begin{figure}[t]
\centering
\includegraphics[width=\columnwidth]{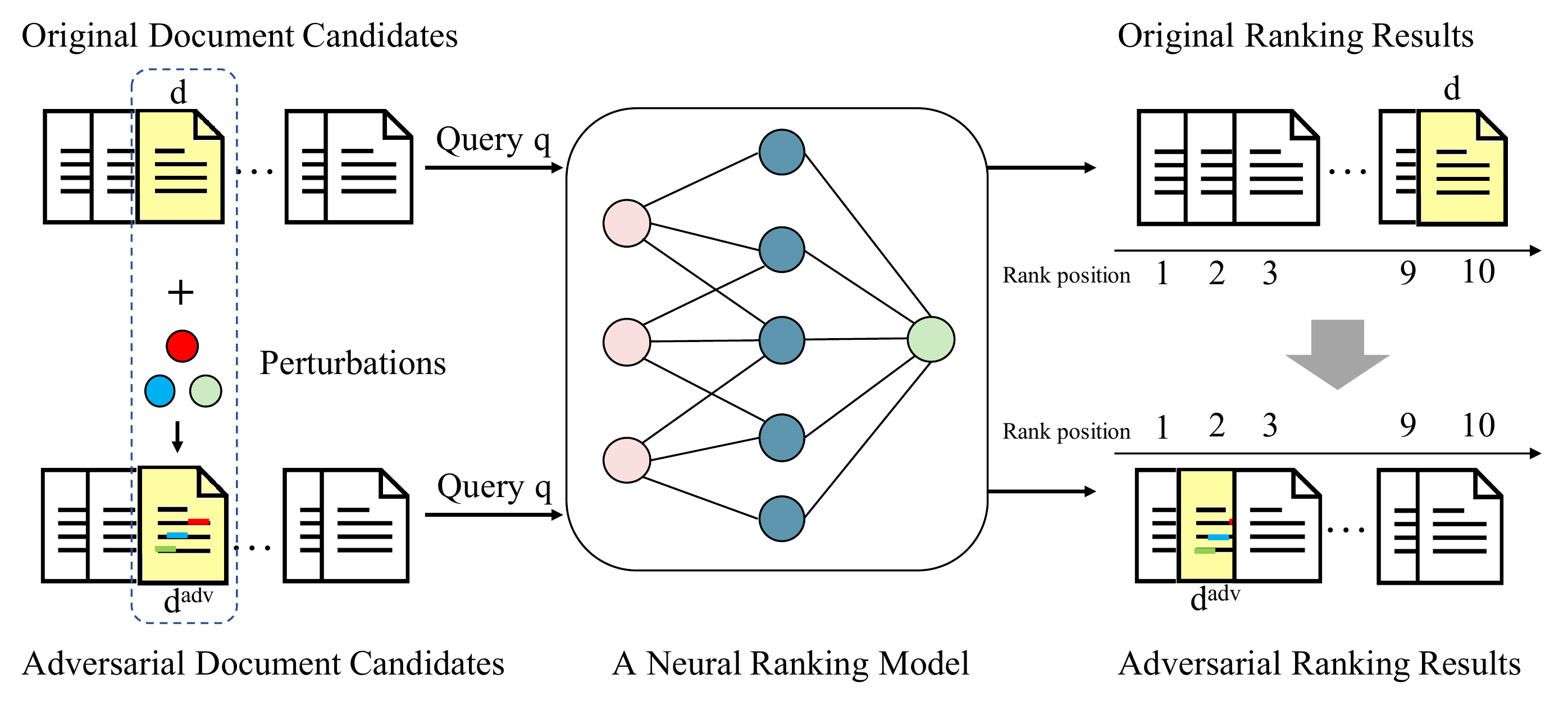}
\caption{Demonstration of the WSRA task. Given a neural ranking model, adversarial perturbation is added to the target document $d$ and the adversarial example $d^{adv}$ will be promoted in rankings with respect to the query $q$.}
\label{fig:ADRA task}
\end{figure}

\header{Adversarial examples}
Deep neural models are notorious for their vulnerabilities to adversarial examples \cite{szegedy2013intriguing, goodfellow2014explaining}. 
For example, \citet{goodfellow2014explaining} show that a panda image, added with imperceptible perturbations, is misclassified as a gibbon by GoogLeNet \cite{googlenet}. 
\citet{liang2017deep} prove that even tiny modifications to a character or a word can fool state-of-the-art deep text classifiers. 
Recent observations have also shown that rankings can rapidly change due to small, almost indiscernible changes of documents~\cite{goren2018ranking}. 
Hence, adversarial attacks may become a new type of web spamming techniques \cite{web_spam} in the neural network based methods which gain importance in IR. 
Since adversarial examples are maliciously crafted by adding perturbations that are imperceptible to humans to legitimate examples, they may not be detected by traditional anti-spamming methods  \cite{shahzad2021improved}. 
Up to now, little attention has been paid to adversarial attacks against NRMs, except for analyses of the robustness of ranking models carried out by \citet{goren2018ranking}. 
Therefore, we believe it is critical to study potential adversarial attacks to identify the vulnerability of NRMs before they are deployed and help facilitate the development of the corresponding countermeasures.

\header{A new adversarial attack task}
In this paper, we introduce the Word Substitution Ranking Attack (WSRA) task against NRMs.
As shown in Figure \ref{fig:ADRA task}, given a neural ranking model, the WSRA task aims to promote a target document in rankings with respect to the query by replacing important words in its text with their synonyms in a semantic-preserving way.
An effective adversarial sample in WSRA needs to satisfy the following qualities: 
\begin{enumerate*} 
\item imperceptible to human judges yet misleading to NRMs; and 
\item fluent in grammar and semantically consistent with the original document. 
\end{enumerate*}
We discuss the major differences between the WSRA task against NRMs and  adversarial attacks for image retrieval and text classification.
We also define different adversarial settings for the WSRA task in terms of the information that attackers rely on, including \emph{white-box} attacks and \emph{black-box} attacks. 
The black-box attacks are further divided into \emph{score-based} attacks and \emph{decision-based} attacks. For the evaluation of the WSRA task, we define the Success Rate (SR) metric for the attacking and adapt the Perturbation Percentage (PP) and Semantic Similarity (SS) from natural language processing (NLP) for automatic evaluation. 

In this work, we focus on the \emph{decision-based black-box attack setting} for the WSRA task. 
This attack scenario is realistic and important, because most of the real-world search engines are black-boxes and only provide hard-label outputs. 
It is also challenging since the gradient cannot be directly computed and the predicted probability is not provided.

\header{An adversarial ranking attack method}
We make the first attempt to address the WSRA task under the decision-based black-box attack setting. 
Specifically, we introduce a novel Pseudo-Relevance based ADversarial ranking Attack method, or PRADA for short, to generate adversarial samples. 
The key idea is to learn a surrogate model to imitate the behaviors of the target NRM for finding the adversarial perturbations. 
Inspired by the Pseudo Relevance Feedback idea ~\citep[PRF,][]{nrm_weak_supervision} in IR, we query the target NRM and take the top-ranked results as relevant examples to learn a surrogate ranking model. 
Then, we identify the important words in a document which have a high  influence on the final ranking result via the prior-guided gradients  generated by the surrogate model. 
With the important words, we apply Projected Gradient Descent ~\citep[PGD,][]{PGD} to generate gradient-based adversarial perturbations to the embedding space according to the expected ranking order.
Finally, we replace the important word with its synonyms in a semantic-preserving way and repeat this process by iterating the importance words list to find the final adversarial sample.

\header{Experiments}
We conduct experiments on two web search benchmark datasets, the MS MARCO document ranking dataset and the MS MARCO passage ranking dataset. 
We compare with several state-of-the-art adversarial attack strategies and our experimental results show that PRADA can successfully promote the target document in rankings with the highest attack success. 
At the same time, the perturbation percentage is considerably lower than for competing attack methods while the semantic similarity score is comparably high. 

\header{Main contributions}
The main contributions of this paper are 
\begin{enumerate*}
\item We introduce a new WSRA task against NRMs for identifying the vulnerability of NRMs and consequently contributing to the design of robust NRMs;  
\item We make the first attempt to address the WSRA task under the decision-based black-box attack setting, and propose a novel PRADA method based on PRF to generate adversarial examples; and
\item We conduct rigorous experiments to demonstrate the effectiveness of our proposed model.
\end{enumerate*}

\section{related work}

In this section, we briefly review three lines of related work, web spamming, text ranking models and adversarial attacks. 

\subsection{Web Spamming}

Web spamming refers to the actions of manipulating web pages intended to mislead search engines into ranking some pages higher than they deserve \cite{web_spam}.
The consequences of web spamming are that the quality of search results decreases and search engine indexes are inflated with useless pages, which increases the cost of each processed query. 

Existing spamming techniques can be divided into \emph{term spamming} and \emph{link spamming} \cite{web_spam}. 
Term spamming refers to techniques that tailor the contents of a web page's text field (e.g., document body, title, meta tags), in order to make spam pages relevant for some queries \cite{castillo2011adversarial}.  
Link spamming creates link structures that are meant to increase the importance of one or more of their pages \cite{link_spamming}.  
To combat such manipulation, prior work studied the detection of web spamming from the perspective of content analysis \cite{ntoulas2006detecting,piskorski2008exploring} and link analysis \cite{wei2012fighting,benczur2005spamrank,gyongyi2004combating}, respectively.
The proposed WSRA task can be viewed as a new type of web spamming against NRMs.

\subsection{Text Ranking Models}

Information retrieval is a core task in many real-world applications, e.g., web search and digital libraries. 
Ranking models lie at the heart of research on IR. During the past decades, different techniques have been proposed for constructing ranking models, from traditional heuristic methods \cite{salton1975vector}, probabilistic methods \cite{BM25, QL}, to modern machine learning methods \cite{liu2011learning,li2014learning}. 
With the advance of deep learning technology, we have witnessed substantial growth of interest in NRMs \cite{drmm,Duet,dai2019deeper,onal-neural-2018}, achieving promising results. 
Recently, pre-trained models such as BERT \cite{BERT} have been widely adopted for text ranking, showing remarkable success in effectiveness \cite{ma2021b}.
In this paper, we use BERT as the target ranking model to evaluate the attack effectiveness. 
Since neural networks become ever more sophisticated, it is costly to obtain massive amounts of annotated training data.
\citet{nrm_weak_supervision} proposed to address this problem by taking advantage of existing unsupervised methods such as BM25 \cite{BM25} for constructing a weakly annotated training set.
Besides, \citet{izsak2014search} also studied how can a search engine with a relatively weak relevance ranking function compete with a search engine with a much stronger relevance ranking function.
Inspired by the idea, we propose to train the surrogate ranking model with weak supervision signals generated by the target model. 

\subsection{Adversarial Attacks}

Adversarial attacks aim to find a minimal perturbation that maximizes the model's risk of making wrong predictions. 
In the white-box attack setting, attackers have complete access to the target model.
Early researchers \cite{szegedy2013intriguing,goodfellow2014explaining,PGD} have extensively studied adversarial attacks for continuous data, e.g., images. 
For example, the Fast Gradient Sign Method \citep[FGSM,][]{goodfellow2014explaining} utilized the error function of the model output and the target category to generate the adversarial perturbation.
Moreover, Projected Gradient Descent~\citep[PGD,][]{PGD} is an iterative version of FGSM, which is regarded as one of the most powerful attacks \cite{athalye2018robustness}. 

In the black-box attack setting, attackers only have access to the outputs of the target model \cite{HSJA}.
Prior work has explored the black-box attack for many NLP tasks, including text classification \cite{gao2018black,liang2017deep}, sentiment analysis \cite{liang2017deep,alzantot2018generating,textbugger}, and natural language inference \cite{alzantot2018generating,minervini2018adversarially}.
Adversarial attacks for text are challenging due to the discrete input space.
To alleviate the problem, \citet{goodfellow2014explaining} adopted FGSM to generate perturbations in the word embedding space and utilized nearest neighbor search to find the closest words. 
However, such methods treat all words as equally vulnerable and replace them with their nearest neighbors, which leads to non-sensical and word-salad outputs \cite{xu2021grey}. 
To tackle the problem, a number of publications \cite{jia2017adversarial,textfooler,liang2017deep}
have adopted heuristic rules to find important words and substitute these words with synonyms. 
Besides, \citet{raval2020one} explored to lower the rank of a document by token changes.
Recently, \citet{goren2020ranking} proposed to promote the rank of the document by replacing a passage in the document with some other passage. However, their evaluation for content-quality maintenance relies on the human judges, and their study is conducted on feature-based learning to rank models. 
In this work, we propose more automatic evaluation metrics for convenient comparison and study prevalent neural ranking models.

Adversarial attacks have also been widely studied in the context of image retrieval. 
For example, \citet{yang2018adversarial} degraded the ranking quality by maximizing the Hamming distance to its own embedding.
\citet{DAIR} proposed a query-based black-box attack against image retrieval models to subvert the top-$k$ retrieval results. 
\citet{adv_rank_attack_and_defense} designed a triplet-like objective function, and combined it with PGD to efficiently obtain the desired adversarial perturbation. 
In this work, we adopt the PGD to perturb the embedding space according to the expected ranking order. 



\section{Problem Statement}

In this section, we introduce the Word Substitution Ranking Attack (WSRA) task against NRMs, and then describe different adversarial attack settings for the WSRA task.  

\subsection{Task Description}

Typically, given a query $q$ and a set of document candidates $ \mathcal{D} = \{d_1, d_2, \ldots, d_N \} $ selected from a document collection $\mathcal{C}$ ($\mathcal{D} \subseteq \mathcal{C}$),
a ranking model $f$ aims to predict the relevance score $\{f(q,d_n)|n=1,2,\ldots,N\}$ between every pair of query and candidate document for ranking the whole candidate set. 
For example, the ranking model outputs the ranked list $L = [d_N, d_{N-1},\ldots, d_1]$ if it determines $f(q,d_N) > f(q,d_{N-1}) \cdots > f(q,d_1)$.

Based on these, the WSRA task aims to fool the NRMs to promote a target document in rankings by replacing important words in its text with their synonyms in a semantic-preserving way. In particular, we assume that the attacker is inclined to select $\mathcal{D}$ from the top ranked documents, as the ranked lists returned to the clients are usually ``truncated'' (i.e., only the partial top-ranked documents will be shown). 

Given an original target document $d$, the goal of an attack is to generate a valid adversarial example $d^{adv}$ in the vicinity of $d$ that is ranked higher by NRMs. Specifically, $d^{adv}$ is crafted to conform to the following requirements, i.e.,
\begin{equation}
\label{problem_form}
Rank_L(q,d^{adv}) < Rank_L(q,d)  \text{ such that } \operatorname{Sim} (d, d^{adv}) \geq \epsilon, 
\end{equation} 
where the adversarial example $d^{adv}$ can be regarded as $d + p$, and $p$ denotes the perturbation to $d$. Rank$_{L}(q, d)$ and Rank$_{L}(q, d^{adv})$ denote the position of the original $d$ and its adversarial example $d^{adv}$ in the ranked list $L$ with respect to the query $q$, respectively.  
A smaller rank position value represents a higher ranking.  
$\operatorname{Sim}$ refers to the similarity function between the original $d$ and its adversarial example $d^{adv}$, and $\epsilon$ is the minimum similarity.  
In the field of natural language, the universal sentence encoder~\cite[USE,][]{USE} is often leveraged as the similarity function $\operatorname{Sim}$.
USE first maps the two inputs into vector using Transformer encoder, and then computes their cosine similarity as the semantic similarity \cite{textfooler, bert-attack, li2021searching}.


Note we can find clear differences between the WSRA task and  adversarial attacks in image retrieval and text classification: 
\begin{enumerate*}
\item The WSRA task needs to ensure that the perturbed document is semantically consistent with the original document by imposing a semantic similarity constraint, while the attack against image retrieval makes the pixel-level perturbations bounded in the budget; and
\item The WSRA task needs to promote the rank positions in a partial retrieved list, instead of misclassifying the single adversarial sample as in  text classification. 
\end{enumerate*}

Specifically, in this work, we choose the fine-tuned BERT model on downstream search tasks for adversarial ranking attack, due to the following:
\begin{enumerate*}
\item the pre-trained language model BERT has shown good superiority on many text ranking problems \cite{gu2020speaker, joshi2020spanbert, ma2021b,lin-2021-pretrained} in both academia and industry in recent years; and
\item previous studies have shown that it is challenging to adversarially attack a fine-tuned BERT on downstream tasks due to its strong performance.
\end{enumerate*}

\subsection{Attack Setting}

Attacks that cause the neural ranking model to purposefully promote a target document in the ranking come in two kinds:

\begin{description}[leftmargin=\parindent,nosep]

\item[\textbf{White-box:}] 
Under the white-box setting, the target model can be fully accessed by attackers.
The attackers can directly obtain the real gradient of the loss for the gradient-based attack, which is often conducted by optimizing an attack objective function \cite{QAIR}.

\item[\textbf{Black-box:}] 
Compared with the white-box attack, the black-box attack is more realistic, since no model information (e.g., parameters and gradients) is available for attackers in reality. 
The attackers can only query the target model to achieve the corresponding output.
Generally, we can divide black-box attacks into score-based attacks and decision-based attacks.

\begin{description}[leftmargin=\parindent]
\item[\textbf{Score-based}:] 
``Score-based'' means that the attacker could leverage the relevance score of each candidate document with respect to the query to conduct the attack.
\item[\textbf{Decision-based}:] 
While attackers can still obtain the relevance score under the score-based setting, only the final decision (i.e., rank positions of the partially retrieved list) could be accessed by attackers under the decision-based setting.
Therefore, the decision-based setting is more challenging.
\end{description}
\end{description}

\begin{figure*}[t]
\centering
\includegraphics[scale=0.35]{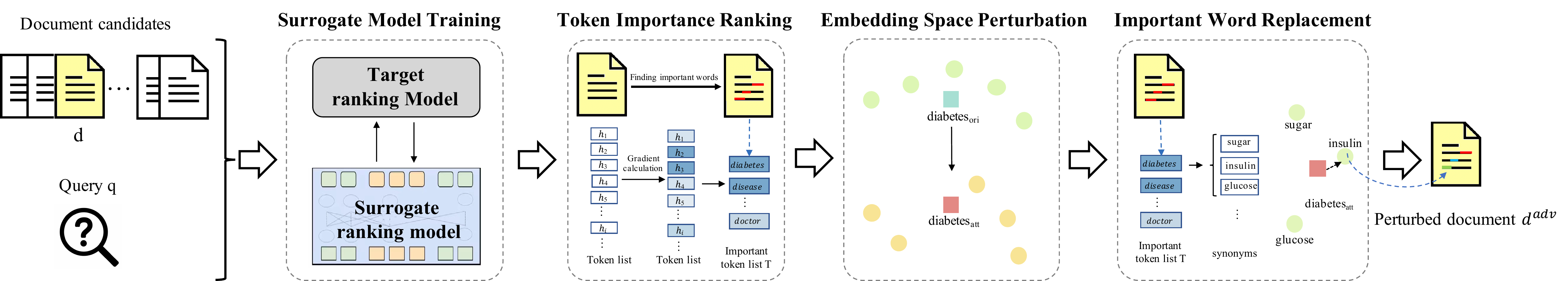}
\caption{The overall architecture of the PRADA method. We first query the target NRM and learn a surrogate ranking model based on PRF idea. Then, we select the important words in a document based on the surrogate model. We apply PGD to generate gradient-based adversarial perturbations to the embedding space towards the expected ranking order. Finally, we iteratively replace the important words with synonyms to find the final adversarial sample.}
\label{fig:overview}
\end{figure*}

\noindent%
In this work, we focus on the \emph{decision-based black-box attack setting} for the WSRA task. 
Although this setting is significantly more challenging than white-box and score-based black-box attacks to NRMs, it is more practical and enables to apply our methods to attack a real-world search engine.


\subsection{Evaluation Metrics}
\label{task_evaluation}
For the evaluation of the WSRA task, we set up the Success Rate (SR) metric for the document ranking attacking, i.e., 
\begin{description}[leftmargin=\parindent,nosep]
\item[\textbf{Success Rate (SR)}] evaluates the percentage of the after-attack documents that are ranked higher than original documents. We define SR as 
   $$
    \text{SR} = \frac{1}{|Q|}\sum_{t=1}^{|Q|}\frac{1}{N_{q}}\sum_{i=1}^{N_{q}}\mathbb{I}\{Rank_{L}(d_i + p) < Rank_{L}(d_i)\},
    $$
    where $|Q|$ denotes the number of evaluated queries, $N_{q}$ the number of attacked documents with respect to each query, and $d_i$ the attacked document with respect to the query $q \in Q$.
    $\mathbb{I}\{\cdot\}$ is the indicator function.
    The effectiveness of an adversarial attack is better with a higher SR (\%). 
\end{description}
Furthermore, we adapt the Perturbation Percentage (PP) and Semantic Similarity (SS) from NLP to measure the quality of the generated samples:
\begin{description}[leftmargin=\parindent,nosep]
    \item[\textbf{Perturbation Percentage (PP)}] evaluates the word-level perturbation percentage of candidate documents following~\cite{bert-attack}. 
    A lower PP (\%) results in higher semantic consistency. 

    \item[\textbf{Semantic Similarity (SS)}] evaluates the semantic similarity between the original document and the adversarial example. 
   Following ~\cite{textbugger, textfooler}, we use the USE to measure the semantic similarity. We set the encoding model to the released deep averaging network\footnote{https://tfhub.dev/google/universal-sentence-encoder/2} since it can encode long documents quickly.
   In this work, we evaluate SS at both the document-level (\textbf{SS$_{doc}$}) and sentence-level (\textbf{SS$_{sen}$}). 
   For SS$_{doc}$, we directly input two documents and evaluate the semantic similarity between them.
   For SS$_{sen}$, we first split two documents into sentence pairs and then evaluate the average sentence semantic similarity between these sentence pairs.
   A higher SS (\%) results in higher semantic consistency. 

\end{description}

\begin{algorithm}[t]
\small
 \caption{PRADA}
 \label{alg}
 \begin{algorithmic}[1] 
 \REQUIRE a query $q$, a pre-collected query collection $Q_C$, a set of document candidates $\mathcal{D}$, a target ranking model $f$, a target document $d$
 \ENSURE an adversarial document $d^{adv}$ \\
 \STATE \textbf{Procedure} Surrogate Model Training 
 \FOR{$q_c \in$ $Q_C$}
 \STATE Get the ranked list $ \boldsymbol{L}_c $ by querying the target model with $q_c$.
 \ENDFOR
 \STATE Train the surrogate model $f_s$ in terms of Eq.(\ref{eq:surrogate_model_training}).
 \STATE \textbf{Procedure} Token Importance Ranking  
 \STATE $H = \{h_1, h_2, ..., h_i, ... \}$ // sub-word token list of $d$
 \STATE Compute the importance score $I_{h_i}$ for each $h_i$ in terms of Eq.(\ref{eq:importance score}).
 \STATE Rank $H$ in descending order to create $T[:m]$ 
 \STATE \textbf{Procedure} Embedding Space Perturbation 
 \FOR{$t\leftarrow 1$ to $\eta$} 
 \STATE Conpute the gradient $\boldsymbol{g}_{d^{adv}_t}$ of Eq. (\ref{loss_RA}) using Eq.(\ref{eq:gradient})
 \STATE Update the adversarial candidate $d^{adv}_{t+1}$ in terms of Eq.(\ref{upatecandidate})
 \ENDFOR
  \STATE Obtain the perturbed vectors $\textbf{o}^p$ of the $m$ important tokens 
 \STATE \textbf{procedure} Important Word Replacement
 \STATE Initialization: $d^{opt} \leftarrow d$
 \FOR{$o_i \in T[:m]$ }
 \STATE Find the corresponding whole word $w_{o_i}$,  $\boldsymbol{e}_{cf}(w_{o_i}) \leftarrow$ map($w_{o_i}$) 
 \STATE Obtain the $S$ synonyms $\{w_s \}_{s=1}^S$ in terms of Eq.(\ref{eq:find_synonyms})
 \STATE $\textbf{e}_{w_s} \leftarrow $ encode($w_s$)
 \STATE $w_s* = \argmax_{w_s\in\{w_s\}_{s=1}^S} \text{CosSim}(\textbf{e}_{w_s}, \boldsymbol{e}_{cf}(w_{o_i}) )$
 \IF{$Rank_{L}(q, d^{temp}) < Rank_{L}(q, d^{opt})$}
 \STATE $d^{opt} \leftarrow d^{temp}$
 \ENDIF
 \ENDFOR
 \RETURN $d^{adv} = d^{opt}$
 \end{algorithmic}
\end{algorithm}

\section{Our Attack Method}

In this section, we introduce our proposed attack method for the WSRA task under the decision-based black-box attack setting. 
We first give an overview of the model architecture and then describe each component of the model in detail.

\subsection{Model Overview}

In this work, we formulate the attack goal of the WSRA task that promotes the target document $d$ in rankings with respect to a query $q \in  Q=\{q_1,\dots,q_{|Q|}\}$ by perturbation $p$ as the following problem: 
\begin{equation}
\label{equ:perturbation}
    p = \arg \min Rank_{L}(q, d + p).
\end{equation}
The optimization problem cannot be directly solved due to the discrete nature of the rank position Rank$_{L}(q, d)$. 
To solve the problem, we design a surrogate objective function following \cite{adv_rank_attack_and_defense}. 
The attacking goal in Eq. (\ref{equ:perturbation}) can be converted into a series of inequalities, i.e., 
\begin{equation}
\label{eq:inequality}
    Rank_{L}(q, d+p) < Rank_{L}(q, L_{\backslash d}),
\end{equation}
where $d$ and $L_{\backslash d}$ denote the target document and the remaining documents from the ranked list $L$, respectively. 
Each inequality represents a pairwise ranking sub-problem between $d$ and other documents $L_{\backslash d}$. 
The adversarial candidate $d^{adv}=d+p$ should be ranked ahead of other documents with respect to $q$.  

Here, we leverage the pairwise hinge loss to model the expected ranking order, i.e., 
\begin{equation}
\label{loss_RA}
    L_{RA}(q, d+p; L) = \sum_{d' \in L_{\backslash d}} \max(0, \beta - f_s(q, d + p) + f_s(q, d') ),
\end{equation}
where $\beta$ is the margin for the hinge loss function, which is often set to 1, and $f_s$ denotes the relevance score given by the surrogate ranking model, which will be described next; $d'$ denotes the remaining documents in the ranked list $L$ without the target document $d$.

In this way, the original problem in Eq.~(\ref{equ:perturbation}) can be reformulated into the following optimization problem:
\begin{equation}
\label{final_opt}
    p = \arg \min L_{RA}(q, d+p; L).
\end{equation}

To ensure the quality of the adversarial examples that are being generated, we further impose constraints on the ranking attack on the following three aspects:
\begin{enumerate*}
\item the maximum number of modified tokens in a document, $m$, 
\item the maximum number of one word's synonyms, $S$, and 
\item the minimum semantic similarity between the original target document and the adversarial example, $\epsilon$. 
\end{enumerate*}

To solve the optimization problem in Eq.~(\ref{final_opt}) and satisfy the required constraints, we introduce a novel Pseudo-Relevance based ADversarial ranking Attack method, or PRADA for short.  
The overall architecture of PRADA is depicted in Figure~\ref{fig:overview}. 
A pseudo algorithm for PRADA is provided in Algorithm \ref{alg}.

Briefly, PRADA can be decomposed into four dependent components: 
\begin{enumerate*}
    \item Surrogate Model Training, to learn a surrogate model that can imitates the behaviors of the target NRM based on the PRF idea; 
    \item Token Importance Ranking, to find the important words in the document that have a  strong influence on the rankings; 
    \item Embedding Space Perturbation, to generate the desired adversarial perturbation in the embedding space for the important words; and
    \item Important Word Replacement, to iteratively replace the important words one by one with synonyms to find adversarial samples that can mislead the target model. 
\end{enumerate*}
Below, we discuss each of the components.

\subsection{Surrogate Model Training}
\label{sec:surrogate model training}

In adversarial attacks, the gradients for guiding the attack process are usually calculated based on knowledge of the target model, which is unavailable under the black-box setting. 
Hence, based on the PRF idea in IR, we propose to train a surrogate ranking model \cite{papernot2016transferability, papernot2017practical} with similar behaviors of the target model. 
Then, we can obtain prior-guided gradients, and attack the target ranking model based on the surrogate model due to the transferability  \cite{papernot2016transferability}. 

\begin{figure}[h]
\centering
\includegraphics[scale=0.25]{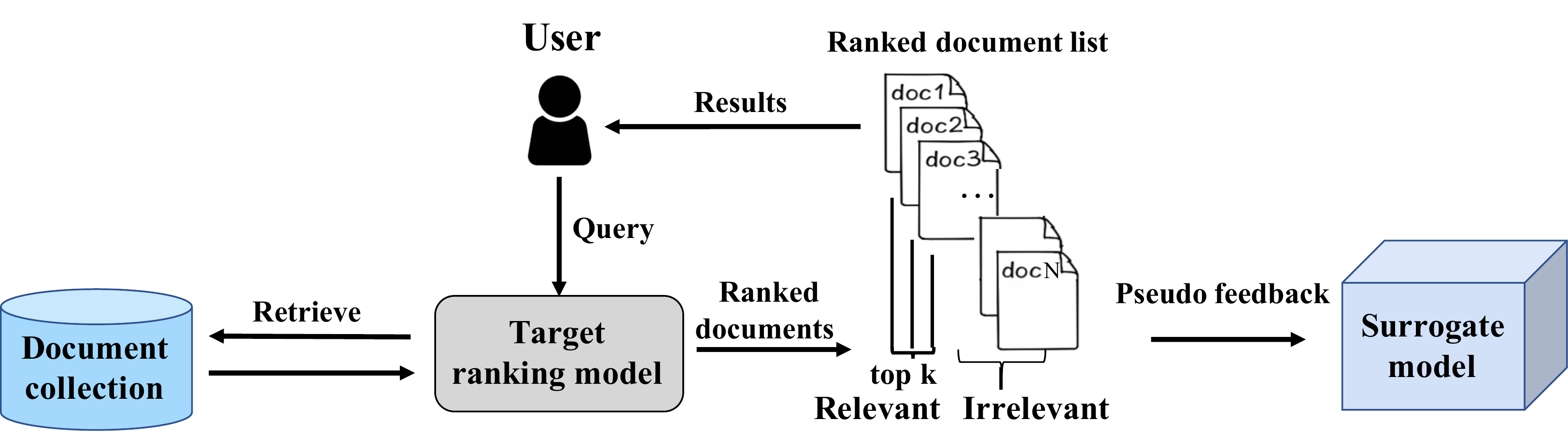}
\caption{The training process for the surrogate model.}
\label{fig:surrogate_model_training}
\end{figure}

\noindent%
Specifically, we leverage the fine-tuned BERT as the target ranking model.  
We initialize the surrogate ranking model using the original BERT. As shown in Figure~\ref{fig:surrogate_model_training}, given a random query $q_c$ from a pre-collected query collection $Q_C$, the target model returns a ranked list $L_c$ with $N$ documents. 
To obtain the priors for attacks, we query the target model with all $|Q_C|$ queries collected from the downstream search tasks.
We generate pseudo-labels as the ground-truth by treating the top-$k$ ranked documents as relevant while treating the other documents as irrelevant, for training the surrogate ranking model $f_s$. 
The objective function is defined as 
\begin{equation}
\label{eq:surrogate_model_training}
\begin{split}
 &L_s ={}\\
 &\frac{1}{|Q_C|}\sum_{c=1}^{|Q_C|} \max(0, \beta - f_s(q_c, L_c[:k]) + f_s(q_c, L_c[k+1:N])),
\end{split} 
\end{equation}
where $L_c[:k]$ denotes the top $k$ ranked documents, and $L_c[k+1:N]$ denotes the remaining documents in the list; $\beta$ is the margin for the hinge loss function, which is often set to 1.

\subsection{Token Importance Ranking}
\label{TIR}

Given a target document $d$, which is tokenized into sub-word token list $H = [h_1, h_2, \ldots, h_i, \ldots]$ by BERT, we observe that only some important tokens act as influential signals for the surrogate ranking model $f_s$, echoing the observation in \cite{textfooler} that BERT attends to the statistical cues of some words. 
That is, perturbations over these important tokens can be most beneficial in crafting adversarial samples. 
Therefore, we propose a scoring mechanism to identify the important tokens in a document which have a high impact on the final ranking result.  

Following \citep{textbugger, texttricker}, we first calculate the gradient magnitude with respect to each input unit. 
Then, we sum up the score for each dimension in the embedding space as the token-level importance score. 
Specifically, we introduce a scoring function that determines the importance $I_{h_i}$ of the $i$-th token $h_i$ in $d$ as
\begin{equation}
\label{eq:importance score}
    I_{h_i} = \left\| \frac{\partial L_{RA}}{\partial \boldsymbol{e}_{h_i}^o} \right\|^2_2,
\end{equation}
where $\boldsymbol{e}_{h_i}^o$ is the original embedding vector of $h_i$ in the surrogate model; 
$L_{RA}$ denotes the adversarial ranking objective function, which is defined in Eq.~(\ref{loss_RA}).  

We rank all the tokens according to the importance score $I_{h_i}$ in descending order to create the candidate token list $T$. 
We only attack the top $m$ important tokens for each $d$, i.e., $T[:m] = [o_1, o_2, \ldots, o_m]$, since we intend to keep the perturbation to a minimum. 

\subsection{Embedding Space Perturbation}
\label{ESP}

NRMs usually map samples (i.e., queries and documents) to an embedding space, where the distances among them determine the final ranking order~\citep{adv_rank_attack_and_defense}. 
A document's position in the embedding space may be changed by adding a perturbation to its important tokens.  
Therefore, we generate gradients based on the surrogate model for finding a proper perturbation to the important tokens, which could push the document to a desired position.  

Specifically, we adopt the Projected Gradient Descent \cite[PGD,][]{PGD} method, which is one of the most effective first-order gradient-based algorithms. Note that in this work, the perturbation $p$ is achieved at the token-level instead of at the document-level. 

For all $m$ important tokens $o_i \in T[:m]$, the PGD algorithm alternates two steps at every iteration $t=1,2,\ldots,\eta$:
\begin{itemize}[leftmargin=*,nosep]
	\item Calculating the gradient $\boldsymbol{g}_{d^{adv}_t}$ of Eq. (\ref{loss_RA}), i.e., 
	\begin{equation}
\label{eq:gradient}
    \boldsymbol{g}_{d^{adv}_t} = \frac{\partial L_{RA}(q, d^{adv}_t; L)}{\partial d^{adv}_t}, 
    \end{equation}
where $d^{adv}_t$ denotes the adversarial document with the embedding of all the important tokens perturbed at the $t$-th step.

\item Leveraging the gradient $\boldsymbol{g}_{d^{adv}_t}$ to update the  adversarial candidate, i.e., 
    \begin{equation}
    \label{upatecandidate}
    d^{adv}_{t+1} = d^{adv}_t - \alpha \frac{\boldsymbol{g}_{d^{adv}_t}}{\left\|\boldsymbol{g}_{d^{adv}_t}\right\|^2_2},
\end{equation}
where $\alpha$ denotes a constant hyper-parameter indicating the PGD step size and $d^{adv}_1$ is initialized as the original $d$. Note that we removed the clip operation in the original PGD algorithm since we have found that it limits the perturbation in the embedding space, which leads to poor experimental results. 
\end{itemize}

\noindent%
After $\eta$ iterations for all the important token $o_i$, we obtain the final perturbed vectors of the $m$ important tokens $T[:m]$, i.e., $\textbf{o}^{p}=\{\boldsymbol{e}_{o_1}^{p},\boldsymbol{e}_{o_2}^{p},\dots,\boldsymbol{e}_{o_m}^{p}\}$. 

\subsection{Important Word Replacement} 
\label{IWR}

Based on the perturbed vectors of $m$ important tokens $T[:m]$, we replace the important token with semantically similar and grammatically correct words and repeat this process by iterating the  list $T[:m]$ to find the final adversarial sample. 
Specifically, we generate a set of synonyms for each important token for replacement, to satisfy the requirement of semantic similarity in Eq. (\ref{problem_form}). 

For a target document $d$, the word replacement phase includes the following steps:

\begin{description}[leftmargin=\parindent,nosep]

\item[\textbf{Extracting synonyms for each important token}.] For each important token $o_{i} \in T[:m]$, we first find its corresponding whole word $w_{o_i}$. 
    If $o_i$ is a single word, the corresponding whole word is itself.
    Otherwise, we search back and forth to recover the corresponding whole word. Then, $w_{o_i}$ is mapped into the counter-fitted word embedding space \cite{counter-fitted} where only synonyms are close to each other, to obtain the word vector $\boldsymbol{e}_{cf}(w_{o_i})$. For each $\boldsymbol{e}_{cf}(w_{o_i})$, we obtain the top $S$ synonyms $\{w_s \}_{s=1}^S$ via 
    \begin{equation}
    \label{eq:find_synonyms}
    	\operatorname{Sim}(\boldsymbol{e}_{cf}(w_{o_i}), \boldsymbol{e}_{cf}(w_s)) \geq  \lambda,
    \end{equation}
    where $\operatorname{Sim}$ denotes the cosine similarity between two counter-fitted embeddings, $w_s$ denotes the synonym of $w_{o_i}$, and $\lambda$ denotes the minimum similarity between $w_{o_i}$ and $w_s$.   
    Furthermore, for each synonym $w_s$ with respect to $w_{o_i}$, we encode it back to the embedding space of the surrogate model to obtain the embedding $\textbf{e}_{w_s}$. Note that if the synonym is tokenized by BERT, $\textbf{e}_{w_s}$ is obtained by the average of sub-word token embeddings.
    
 \item[\textbf{Greedy word replacement strategy}.] We calculate the cosine similarity between the candidate synonym vector $\textbf{e}_{w_s}$ and the corresponding perturbed word vector $\boldsymbol{e}_{o_i}^p \in \textbf{o}^{p}$.  
    The synonym $w_s^*$ which has the highest cosine similarity with $w_{o_i}$ is chosen to replace $w_{o_i}$. Suppose the document before this word replacement process is $d^{opt} = \{w_1, w_2, \ldots, w_{o_i}, \ldots\}$, the document after the word replacement is $d^{temp} = \{w_1, w_2, \ldots, w_{o_{i-1}}, w_s^*, w_{o_{i+1}} \ldots\}$. 
    Simply replacing a token by its synonym cannot guarantee a successful attack. Therefore, we adopt a greedy word replacement strategy. Specifically, we obtain the rank of $d^{temp}$ by querying the target model. 
    If the rank of $d^{temp}$ has improved, i.e., $Rank_{L}(q, d^{temp}) < Rank_{L}(q, d^{opt}),$
    we accept the replacement and denote $d^{opt}$ as the $d^{temp}$, i.e., 
    $d^{opt} \leftarrow d^{temp}.$ 
    Otherwise, we will discard this word replacement and turn to the next round.
    
\end{description}

\noindent%
The process described above is repeated by iterating over the importance word list $T[:m]$ to find the final adversarial sample $d^{adv}$.

\section{Experimental Setup}

In this section, we introduce our experimental settings. 


\subsection{Datasets}
To evaluate the performance of our model, we conducted experiments on two web search benchmark datasets. 
\begin{itemize}[leftmargin=*,nosep]
    \item \textbf{MS MARCO Document Ranking dataset} \cite{nguyen2016ms} (MS-MARCO-Doc) is a large-scale benchmark dataset for web document retrieval, with about 3.21 million web documents.  
    \item \textbf{MS MARCO Passage Ranking dataset} \cite{nguyen2016ms} (MS-MARCO-Pas) is a large-scale benchmark dataset for passage retrieval, with about 8.84 million passages from web pages. 
  \end{itemize}

\noindent%
Detailed dataset statistics are shown in Table~\ref{table:dataset statistics}. We take these datasets for experiments since 
\begin{enumerate*}
\item Relevant documents for each user's query are retrieved using Bing from its large-scale web index, which is representative of real web search scenario. 
\item It is practical to promote irrelevant documents instead of relevant documents in rankings. 
The probability of selecting relevant document for attack is low since each query has only one relevant document.  
\end{enumerate*}

\begin{table}[t]
\centering 
 \setlength\tabcolsep{1.1pt}
 \caption{Data statistics.  \#w denotes the number of words.}
\begin{tabular}{lcc} 
\toprule 
 & MS-MARCO-Doc  & MS-MARCO-Pas \\
\midrule
Training queries & 0.37M & 0.5M  \\
Dev queries  & 5,193 & 6,980  \\ 
\midrule
Documents/passages & 3.21M & 8.84M  \\
Documents/Passages: avg \#w & 1,129  & 58 \\
\bottomrule
\end{tabular}
\label{table:dataset statistics}
\end{table}

\subsection{Baselines}

We adopt two types of baselines for comparison, including step-wise methods and traditional term spamming methods. 

\subsubsection{Step-wise Methods}
For step-wise methods, we apply two steps to attack the target document, where the first step is to select $n$ words in the document, and the second step is to substitute these words. 
For the word selection step, we employ four methods:
\begin{itemize}[leftmargin=*,nosep]
    \item \textbf{First} selects the first $n$ words in the document to attack.
    \item \textbf{Last} selects the last $n$ words in the document to attack.
    \item \textbf{Tf-idf} selects the top $n$ words with the highest tf-idf scores in the document to attack.
    \item \textbf{TextRank} selects $n$ words by TextRank \cite{mihalcea2004textrank}, a graph-based method inspired by the PageRank algorithm.
\end{itemize}

\noindent%
For the word replacement step, we employ two methods:
\begin{itemize}[leftmargin=*,nosep]
    \item \textbf{Random Replacement (RR)} replaces the selected word with a random word. 
    \item \textbf{Nearest Replacement (NR)} replaces the selected word with the nearest word in the Glove \cite{pennington2014glove} using cosine similarity.  
   \end{itemize}

\noindent%
By combining these two-step methods, we obtain eight types of attack methods denoted as \textbf{First+RR}, \textbf{First+NR}, \textbf{Last+RR}, \textbf{Last+NR}, \textbf{Tf-idf+RR}, \textbf{Tf-idf+NR}, \textbf{TextRank+RR}, and \textbf{TextRank+NR}.

\subsubsection{Traditional Term Spamming Methods}

Term spamming \cite{web_spam} refers to techniques that tailor the contents of a web page's text fields to rank it higher than they deserve.
Here, we apply two traditional term spamming methods: 

\begin{itemize}[leftmargin=*,nosep]
    \item \textbf{Repetition (TS$_{Rep}$)} promotes the rank of $d$ by adding a small number of query terms~\cite{web_spam}. 
    We randomly choose a starting position in $d$ and replace the following successive $n$ words with $n$ query terms. 
    \item \textbf{Stitching (TS$_{Sti}$)} is to manually glue together sentences from other documents~\cite{web_spam}. We randomly choose a starting position in $d$ and replace the following successive $n$ words with $n$ words extracted from a  sentence pool $S_{pool}$. 
    Following \cite{goren2020ranking} where authors tend to mimic content in documents that were highly ranked in the past for a query of interest, we construct $S_{pool}$ by collecting sentences in documents that are ranked higher than $d$. 
\end{itemize}

\subsection{Model Variants}

We implement several variants of PRADA by removing major components, and adopting different strategies:

\begin{itemize}[leftmargin=*,nosep]
    \item \textbf{PRADA$_{-TIR}$} removes the step of finding important words described in Section \ref{TIR}, and randomly selects words to attack. 
    \item \textbf{PRADA$_{-ESP}$} removes the embedding space perturbation described in Section \ref{ESP}, and applies random perturbations on the embedding space. 
    \item \textbf{PRADA$_{-IWR}$} removes the important word replacement described in Section \ref{IWR}, and replaces all important words in the document with words that are nearest to the perturbed word vectors.
    \item \textbf{PRADA$_{-ESP-IWR}$} removes both the embedding space perturbation and word replacement. It applies random perturbations on the embedding space and directly selects the nearest word to replace the important word.
\end{itemize}

\subsection{Evaluation Metrics}
\label{sec:evaluation metrics}

Besides the automatic evaluation metrics defined in Section \ref{task_evaluation}, we further conduct human evaluations to measure the  quality of the attacked documents from three aspects: 
    \begin{enumerate*}
    \item fluency in grammar; 
    \item imperceptibility to human judges; and 
    \item semantically consistency with original documents. 
    \end{enumerate*}

We first randomly sample 40 test queries from MS-MARCO-Doc and take the corresponding 9 original documents for each query. 
    Then, we find the 360 adversarial samples generated by PRADA and TR$_{rep}$, respectively. 
    We shuffle a mix of original and adversarial documents (i.e., 1,080 in total) and asked three labelers to evaluate them. 
    For (1), annotators score the quality of the mixed examples from 1--5 following \cite{bert-attack}. 
    For (2), annotators judge each example whether it is attacked (i.e., labeled as 0) or not (i.e., labeled as 1).
    For (3), we compare adversarial samples generated by PRADA and TR$_{rep}$ with the original documents, using the following criteria: i) 2: the adversarial sample is completely semantically consistent with the original document; ii) 1: the adversarial sample is partially relevant with the original document and human can still understand the original information; and iii) 0: the adversarial sample is not relevant with the original document. 
    Agreements to measure inter-rater consistency among three labelers are calculated with the Fleiss' kappa \cite{fleiss1973equivalence}.

\subsection{Implementation Details}

In the surrogate ranking training process, the target model is obtained by fine-tuning BERT on the training queries of the MS-MARCO-Doc and MS-MARCO-Pas, respectively.

To train the surrogate ranking model, we leverage the test queries of the MS-MARCO-Doc and MS-MARCO-Pas as $Q_c$, respectively.  
Following \cite{dai2019deeper}, we apply BERT$_{base}$ released by Google.   
For the MS-MARCO-Doc, we use the official top 100 ranked documents retrieved by the QL model following~\cite{craswell2020overview}. 
For the MS-MARCO-Pas, initial retrieval is performed using the Anserini toolkit~\cite{yang2018anserini} with the BM25 model to obtain the top 100 ranked passages following \cite{ma2021prop}. 
The ranked list $L_c$ is obtained by utilizing the target BERT model to re-rank the above initial ranked list and the length $N$ is set to 100.  
We set $k=1$ in Eq.~(\ref{eq:surrogate_model_training}) since every query in the MS-MARCO-Doc and most queries in the MS-MARCO-Pas have only one relevant document. 
In the token importance ranking process, the number of top important tokens $m$ in PRADA is set to 50 and 20 for the MS-MARCO-Doc and MS-MARCO-pas, respectively. 
For fair comparison with the baselines, we also set $n$ to 50 and 20 for the MS-MARCO-Doc and MS-MARCO-pas, respectively.  
Besides, we will analyze the effect of $m$ in PRADA on the attack performance.  
In the embedding space perturbation process, the PGD step size $\alpha$ is set to 45 and the number of iteration $\eta$ is set to 3. 
In the important word replacement process, we set the minimum similarity $\lambda$ to $0.5$.  
 
We evaluate PRADA on 200 queries (i.e., $|Q|=200$) randomly sampled from the dev set in the MS-MARCO-Doc and MS-MARCO-Pas datasets, respectively. 
For each query, we attack 9 target documents in the top 100 documents, which are obtained by picking 1 out of every 10 documents.
Specifically, we randomly choose 1 document from 9 ranges in the document list, i.e., {[11, 20], [21, 30], \ldots, [91, 100]}, respectively. 
Note that we do not choose documents from the range of [1,10] since it is not necessary to attack the top-10 documents for ranking promotion. 
Code will be available at URL.

\section{Experimental Results}
In this section, we report and analyze the experimental results to demonstrate the effectiveness of the proposed PRADA method.
Specifically, we target the following research questions: 
\begin{enumerate*}[label=(RQ\arabic*)]
\item How does PRADA perform compared with baselines under the automatic and human evaluations?
\item Can PRADA evade the detection by the anti-spamming method? 
\item How do different components of the PRADA affect the performance? 
\item How does PRADA perform for different rank positions in the document list? 
\item How does the number of important tokens $m$ affect the PRADA  performance?
\end{enumerate*}

\begin{table}[t]
\centering
   \caption{Comparisons between PRADA and the baselines under the automatic evaluation; * indicates significant improvements over the next-best approach (p-value < 0.05).}
 \renewcommand{\arraystretch}{1.3}
 \setlength\tabcolsep{2.5pt}
  	\begin{tabular}{l @{~} c c c c   c c c c  }
  \toprule
  \multirow{2}{*}{Method} & \multicolumn{4}{c}{MS-MARCO-Doc} & \multicolumn{4}{c}{MS-MARCO-Pas}  \\ \cmidrule(r){2-5} \cmidrule(r){6-9}
       & SR & PP & SS$_{doc}$ & SS$_{sen}$ & SR & PP & SS$_{doc}$ & SS$_{sen}$ \\ 
       \midrule
First+RR  & 65.9 & 13.0 & 90.9 & 92.0 & 9.3 & 24.5 & 78.1 & 79.7\\
First+NR  & 41.9  & 13.0 & 94.3 & 94.9  & 14.8 & 24.5 & 85.9 & 86.3\\ 
Last+RR   & 10.7& 13.0 & 91.1 & 91.9 & 20.7 & 24.5 & 78.7 & 81.5\\
Last+NR  &  8.2 & 13.0 & 94.7 & 95.1 & 22.7 & 24.5 & 86.2 & 87.1\\
Tf-idf+RR    &  48.1 & 13.0 & 90.5 & 90.5 & 8.8 & 24.5 & 80.5 & 80.5\\
Tf-idf+NR    &  43.8 & 13.0 & 93.1 & 93.0 & 10.1 & 24.5 & 81.5 & 81.1\\
TextRank+RR & 55.4 & 13.0 & 87.4 & 88.8 & 8.7 & 24.5 & 74.2 & 73.7 \\
TextRank+NR & 37.5 & 13.0 & 90.8 & 92.7 & 13.9 & 24.5 & 84.2 & 83.6\\ 
\midrule
TS$_{rep}$ & 93.1 & 12.8 & 87.9 & 89.1 & \textbf{99.5} & 24.0 & 85.6 & 87.5 \\ 
TS$_{sti}$ & 70.9 & 12.9 & 91.2 & 91.7 & 59.9 & 24.3 & 86.8 & 87.0 \\
\midrule
PRADA & \textbf{96.7}\rlap{$^*$} & \textbf{4.0}\rlap{$^*$} & \textbf{95.2}\rlap{$^*$} & \textbf{96.2}\rlap{$^*$} & 91.4 & \textbf{7.8}\rlap{$^*$}& \textbf{93.2}\rlap{$^*$} & \textbf{93.1}\rlap{$^*$} \\
\bottomrule
    \end{tabular}

   \label{table:Baseline}
\end{table}

\subsection{Baseline Comparison} 
\label{sec:baseline}

To answer \textbf{RQ1}, we compare PRADA with different baselines under both the automatic evaluations and human evaluations.

\header{Automatic evaluation}
The performance comparisons between our model and the baselines are shown in Table ~\ref{table:Baseline}. 
For the MS-MARCO-Doc, we have the following observations:
\begin{enumerate*}
\item Step-wise methods generally perform worse than term spamming methods and PRADA in terms of SR, indicating that promoting the document in rankings is a non-trivial problem. 
\item For step-wise methods, the methods based on NR perform better than that based on RR in terms of SS$_{doc}$ and SS$_{sen}$. 
\item Term spamming methods perform the best in terms of SR among the baselines. 
TS$_{rep}$ performs better than TS$_{sti}$, indicating that replacing words in a document with the query terms is better than words from other documents.
\item PRADA performs best in terms of all the automatic evaluation metrics. 
That is, PRADA achieves a high success rate while maintaining a minimum perturbation, indicating the perturbation of the important words would be easier to result in ranking promotion from the target model, which is consistent with previous observations \cite{bert-attack}. 
\end{enumerate*}

When we look at the performance of different models on the MS-MARCO-Pas, we find that PRADA performs worse than TS$_{rep}$ in terms of SR when dealing with short texts. 
A potential reason is that there are few options in the passage for  determining  important tokens. 
This result is consistent with the previous observation~\cite{bert-attack}, where the sequence length plays an important role in the high-quality perturbation process and the word replacement would be less reasonable when dealing with extremely short sequences. 
By conducting further analysis, we find that PRADA prefers to keep the original words due to the unsuccessful attack in the greedy word replacement strategy, which contributes to the semantic consistency. 
In future work, we aim to consider a more advanced objective towards both long text and short text for developing robust NRMs.

\begin{table}[t]
\centering 
 \renewcommand{\arraystretch}{1.3}
 \setlength\tabcolsep{1pt}
 \caption{Comparisons between PRADA and TS$_{rep}$ under the human evaluation.}
\begin{tabular}{l @{} c c | c c | c c} 
\toprule 
 & Grammar & kappa & Imperceptibility & kappa & Semantic & kappa \\
\midrule
Original & 3.50 & 0.373 & 0.88 & 0.475 & - & - \\
TS$_{rep}$ & 1.69  & 0.177 & 0.06  & 0.647 & 0.43 & 0.298  \\
PRADA & 3.23  & 0.478 & 0.85 & 0.486 & 1.37 & 0.412 \\
\bottomrule
\end{tabular}
\label{table:human}
\end{table}

\header{Human evaluation} 
Table \ref{table:human} shows the human evaluation results. We can observe that 
\begin{enumerate*}
\item The semantic consistency and language fluency of the adversarial examples generated by PRADA are better than that generated by TS$_{rep}$.  
The adversarial examples generated by PRADA are more imperceptible to human judges than TS$_{rep}$. 
Intuitively, humans can easily identify an attacked document with  multiple successive repetitive words. 
All the human judgement results again demonstrate the effectiveness of our PRADA method. 
\item The kappa values of PRADA for all three aspects are larger than 0.4, considered as ``moderate agreement'' regarding quality of adversarial examples. 
The largest kappa value (i.e., 0.647) is achieved by TS$_{rep}$ for imperceptibility, which seems reasonable since it is easy to reach an agreement on the attacked documents with successive repetitive words.

\end{enumerate*}

\begin{table}[t]
\centering 
 \renewcommand{\arraystretch}{1.4}
 \setlength\tabcolsep{5pt}
 \caption{The detection rate (\%) of PRADA and TS$_{rep}$ via a  representative anti-spamming method; * indicates statistically significant improvements over TS$_{rep}$ (p-value < 0.05).}
\begin{tabular}{l c c c c c c c} 
\toprule 
$\beta$ & 0.080 & 0.075 & 0.070 & 0.065 & 0.060 & 0.055 & 0.050  \\
\hline
TS$_{rep}$ & 81.0 & 85.8 & 90.2 & 93.4 & 96.2 & 98.2 & 99.4 \\
PRADA &  \phantom{0}7.1\rlap{$^*$} & \phantom{0}8.2\rlap{$^*$} & \phantom{0}9.3\rlap{$^*$} & 11.4\rlap{$^*$} & 13.9\rlap{$^*$} & 15.6\rlap{$^*$} & 19.2\rlap{$^*$} \\
\bottomrule
\end{tabular}
\label{table:spam detection}
\end{table}

\renewcommand{\arraystretch}{1.2}
\begin{table}[t]
\centering
   \caption{Model analysis of PRADA under automatic evaluations; * denotes significant degradation w.r.t.\ PRADA (p-value<0.05).}
   \setlength\tabcolsep{1.1pt} 
  	\resizebox{\columnwidth}{!}{%
  	\begin{tabular}{@{}l @{~} c c c c  c c c c  @{}}  \toprule
  \multirow{2}{*}{Method} & \multicolumn{4}{c}{MS-MARCO-Doc} & \multicolumn{4}{c}{MS-MARCO-Pas}  \\ \cmidrule(r){2-5} \cmidrule(r){6-9}
       & SR & PP & SS$_{doc}$ &  SS$_{sen}$ & SR & PP & SS$_{doc}$ &  SS$_{sen}$ \\ 
    \midrule
PRADA$_{-TIR}$ & 86.1\rlap{$^*$} & 4.0 & 94.8 & 95.7 & 87.1\rlap{$^*$} & 7.8 & 92.7 & 89.9\rlap{$^*$}\\ 
PRADA$_{-ESP}$ & 94.7 & 4.0 & 94.6 & 95.3 & 90.2 & 7.8 & 92.6 & 89.8\rlap{$^*$}\\
PRADA$_{-IWR}$ & 39.6\rlap{$^*$} & 4.5 & 92.5 & 94.8 & 47.8\rlap{$^*$} & 8.5 & 82.6\rlap{$^*$} & 86.4\rlap{$^*$} \\
PRADA$_{-ESP-IWR}$ & \phantom{0}5.8\rlap{$^*$} & 4.5 & 92.3 & 94.7 & 10.6\rlap{$^*$} & 8.5 & 82.4\rlap{$^*$} & 86.1\rlap{$^*$} \\
\midrule
PRADA & \textbf{96.7} & \textbf{4.0} & \textbf{95.2}  & \textbf{96.2} & \textbf{91.4} & \textbf{7.8} & \textbf{93.2} & \textbf{93.1}\\
\bottomrule
    \end{tabular}
    }
   \label{table:Model Variants}
\end{table}

\subsection{Spam Detection}
To answer \textbf{RQ2}, we adopt the utility-based term spamicity method \cite{zhou2009osd}, which can online detect whether target pages are spam or not, to detect the adversarial examples generated by PRADA and the best baseline model TS$_{rep}$ on the MS-MARCO-DOC . 
Specifically, if the spamicity score is higher than a utility threshold $\beta$, such example is detected as a spam. The results of detection rates are shown in Table \ref{table:spam detection}.
We have three main observations:
\begin{enumerate*}
\item The detection rate increases with the decrease of the threshold $\beta$.
\item TS$_{rep}$ can be very easily detected under the spam detection algorithm  since it puts many query terms into documents.
\item PRADA outperforms TS$_{rep}$ significantly (p-value < 0.05). It is much easier for PRADA to evade the spam detection (e.g., for $\beta=0.050$, the detection rate of PRADA and TS$_{rep}$ is less than 20\% and over 99\%, respectively). 
\end{enumerate*}

\begin{table*}[t]
\footnotesize
\centering
  \renewcommand{\arraystretch}{1.5} 
   \setlength\tabcolsep{1.0mm} 
   \caption{Adversarial samples generated by TS$_{rep}$ and PRADA on the MS-MARCO-Doc dataset. The perturbed words are marked as blue and red/magenta in the original document and adversarial example, respectively.}
  \begin{tabular}{c l  c}  \toprule 
 Method & Query: ``government does do'' & Rank Position \\ \hline
  \multirow{3}{*}{Original} & \ldots{} what kind of government does japan have \textcolor{blue}{today}? \textcolor{blue}{answered} by the wiki answers community answers. com is making the world better one& \multirow{3}{*}{60} \\
& answer at a time. japan is a \textcolor{blue}{constitutional} monarchy with a parliamentary government. the constitution. \textcolor{blue}{it} awards the vote to all men and &\\
&  women age 20 and older. was this answer \textcolor{blue}{useful}?  what kind of government did japan \textcolor{blue}{have}? japan has the \textcolor{blue}{type} of government like \textcolor{blue}{canada} \ldots  &\\ \hline
\multirow{3}{*}{TS$_{rep}$} & \ldots{}  what kind of government does japan have today? answered by the wiki answers community answers. com is making the world better one& \multirow{3}{*}{38}\\
& answer at a time. japan is a \textcolor{magenta}{does do government does do government does do government does do government does do government does}&\\
& \textcolor{magenta}{do government does do government does do government does do government does do government does do government does do} like canada \ldots{}  &\\ \hline
\multirow{3}{*}{PRADA} & \ldots{} what kind of government does japan have \textcolor{red}{currently}? \textcolor{red}{answer} by the wiki answers community answers. com is making the world better one& \multirow{3}{*}{35} \\
& answer at a time. japan is a \textcolor{red}{constitution} monarchy with a parliamentary government. the constitution. \textcolor{red}{he} awards the vote to all men and&\\
& women age 20 and older. was this answer \textcolor{red}{helpful}?  what kind of government did japan \textcolor{red}{has}? japan has the \textcolor{red}{types} of government like \textcolor{red}{canadian} \ldots &\\ 
\bottomrule
  \end{tabular}
  \label{table:case study}
\end{table*}

\subsection{Model Ablation}
To answer \textbf{RQ3}, we conduct an ablation analysis to investigate the effect of different components in the PRADA. 
Based on Table~\ref{table:Model Variants}, we observe that: 
\begin{enumerate*} 
\item By removing important word replacement, the performance of PRADA$_{-IWR}$ in terms of SR has a significant drop as compared with  PRADA. 
The results indicate that the greedy synonym replacement strategy does help the rank promotion. 
PRADA$_{-ESP}$ has a similar performance with PARDA, which again demonstrates the effectiveness of the word replacement with synonyms. 
\item PRADA$_{-ESP-IWR}$ performs much worse than the PRADA$_{-IWR}$. 
Without the limitation given by the word replacement, the embedding space perturbation has an obvious influence on the results. 
\item By including all the components, PRADA achieves the best performance among the variants in terms of all evaluation metrics. 
\end{enumerate*}

\subsection{Analysis at Different Rank Positions}
To answer \textbf{RQ4}, we analyze the success rate for documents with different rank positions on the MS-MARCO-DOC by PRADA.

\begin{figure}[t]
 \centering
\begin{tabular}{c}
 \includegraphics[clip,trim=4mm 0mm 3mm 0mm,width=0.675\columnwidth]{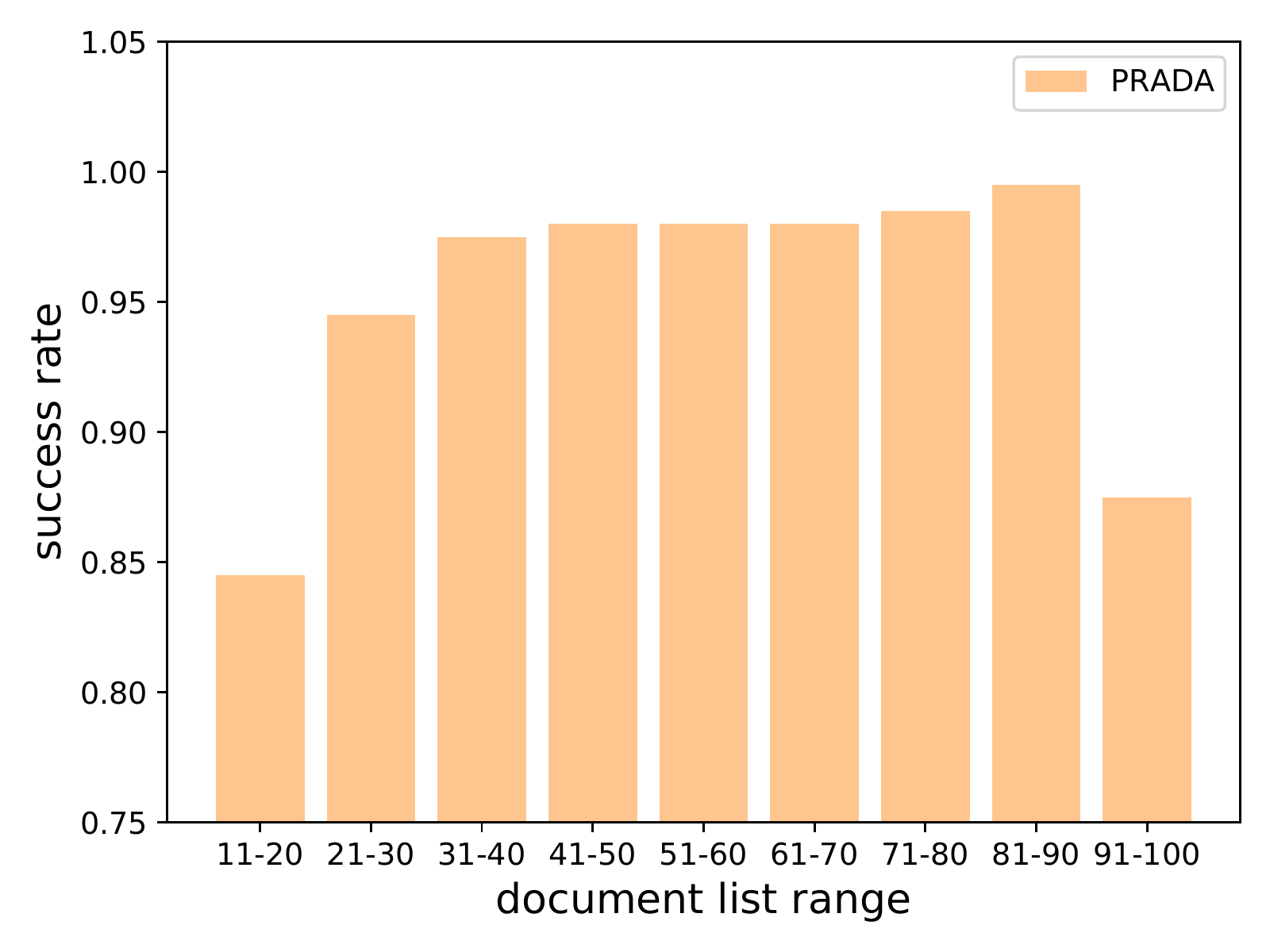}
\\[-1ex]
 \includegraphics[clip,trim=3mm 0mm 16mm 0mm,width=0.675\columnwidth]{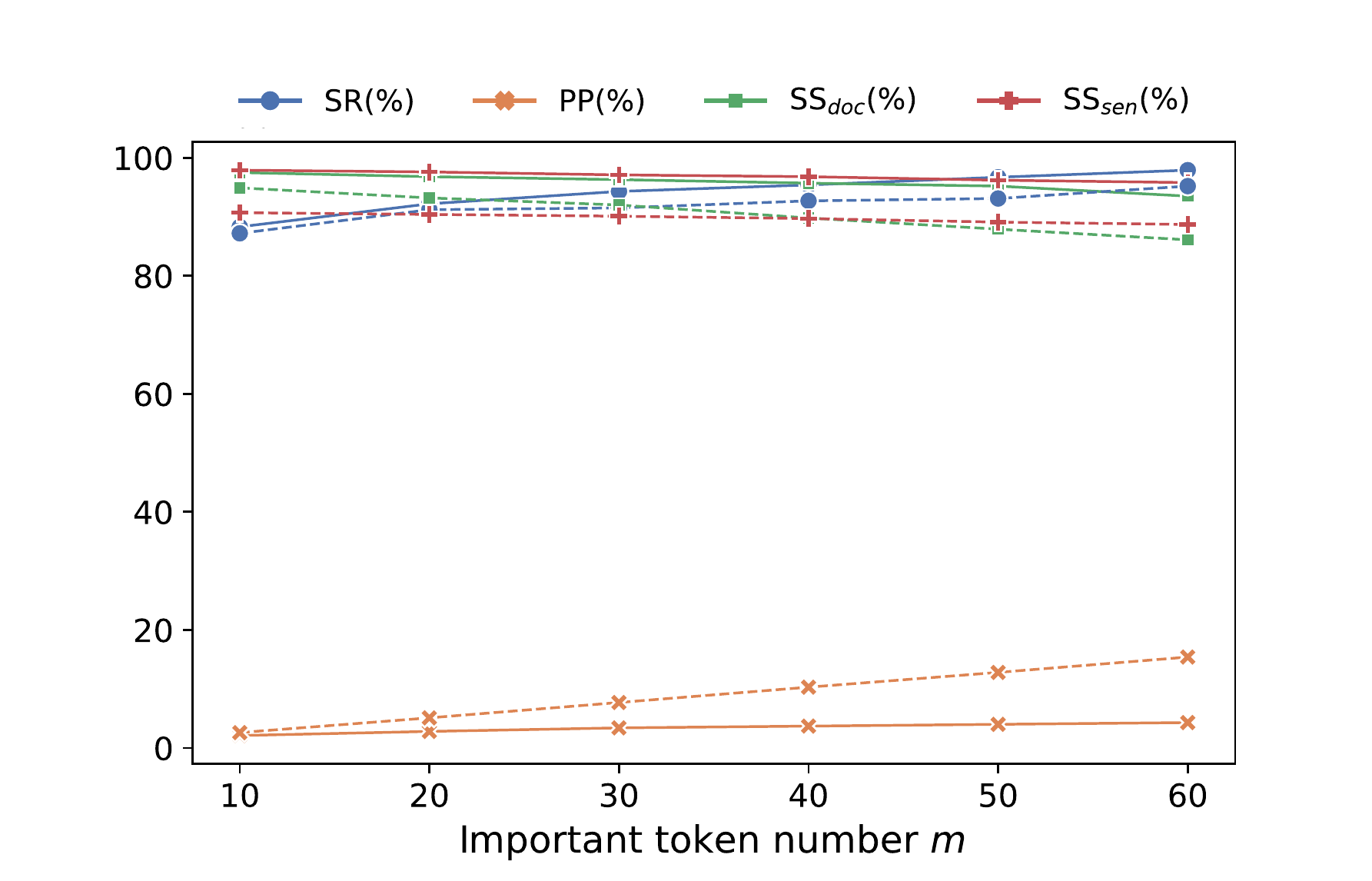}
\\[-1ex]
\end{tabular}
 \caption{(Top): Success rate at different ranges of the ranked list from PRADA on MS-MARCO-Doc. (Bottom): Performance comparison between PRADA and TS$_{rep}$ with different numbers of important tokens on  MS-MARCO-Doc. Dotted lines denote TS$_{rep}$; solid lines denote  PRADA.}
 \label{fig:document_list_range_and_itn}
\end{figure}

Specifically, we visualize the distribution of the success rate in different ranges of the document list (i.e., {[11, 20], [21, 30], \ldots, [91, 100]}) in Figure  \ref{fig:document_list_range_and_itn}(Top).
As we can see: 
\begin{enumerate*}
    \item In general, it is harder to promote high-ranked documents  in rankings than low-ranked documents. Documents in the range of [11,20] are the most difficult to be attacked, with a SR value as  only 0.845. 
    \item It is surprising to find that documents in the last range (i.e., [91,100]) achieve a low success rate (i.e., 0.875). One possible explanation is that these documents are too irrelevant to be promoted in rankings. It is necessary to focus on the attack against low-ranked documents in the future.
\end{enumerate*}

\subsection{Analysis of the Number of Important Tokens} 

To answer \textbf{RQ5},  we analyze the effect of different numbers of important tokens $m$ for PRADA on the attack performance.  
Specifically, we compare PRADA with the best performing baseline TS$_{rep}$ on the MS-MARCO-Doc and set $m$  to six different values (i.e., 10, 20, 30, 40, 50, 60). 
Note that the selected number $n$ of TS$_{rep}$ is equal to $m$. 
As shown in Figure \ref{fig:document_list_range_and_itn}(Bottom), we find that: 
\begin{enumerate*}
\item Overall, the SS and PP increases with the increase of $m$ for both PRADA and TS$_{rep}$. 
This result indicates that attacking more words is more likely to promote the rank.
\item Intuitively, a larger $m$ would result in less semantic similarity. 
The SS$_{doc}$ of TS$_{rep}$ has a larger drop than PRADA in the range of [30,60], and the performance of PRADA in terms of SR and PP is always better than TS$_{rep}$ with different $m$.
These results again illustrate the effectiveness of PRADA.   
\end{enumerate*}

\subsection{Case Study}
To obtain a better qualitative understanding of how different models perform, we show the adversarial examples from PRADA as well as that from  TS$_{rep}$, with the number of important tokens $m$ set to 50. 
We take one query ``government does do'' from the dev set of the MS-MARCO-Doc as an example. Due to space limitations, we only show some key sentences in the document. 
As shown in Table \ref{table:case study}, we can observe that
\begin{enumerate*}
\item Compared with TS$_{rep}$, the adversarial document generated by PRADA is more semantically consistent with the original document by human judges, while the rank position given by the target model is higher (i.e., 35 v.s. 38). 
\item The adversarial document generated by TS$_{rep}$ has a wider range of obvious replacements with query terms, making them distinguishable from the original document and less fluent. Word-level synonyms seem more reasonable for guaranteeing fluency and semantic preservation in  adversarial samples than the query terms. 
\end{enumerate*}

\section{Conclusion and future work}

In this paper, we proposed a challenging WSRA task against NRMs, which aims to promote a target document in rankings by adding adversarial perturbations to its text. 
We focused on the practical decision-based black-box attack setting and 
developed a novel PRADA method based on the PRF idea to generate the adversarial examples for effective attack. 
Empirical results show that the PRADA achieves a high success rate with small indiscernible perturbations. 
Besides, PRADA can evade the detection of the anti-spamming method easily.

A limitation of our PRADA is that the attack may fail the short documents or low-ranked documents. 
In the future work, we aim to go further for stronger black-box attacks against NRMs. 
It is also valuable to attack a real-world search engine by the PRADA to demonstrate its practical applicability.  
We hope this study could provide useful clues for future research on adversarial ranking defense and help develop robust real-world search engines.

\bibliographystyle{ACM-Reference-Format}
\bibliography{references}

\end{document}

%% file: packages.tex
\usepackage{multirow}
\usepackage{graphicx}
\usepackage[inline]{enumitem}
\usepackage{sistyle}
\usepackage{algorithm}
\usepackage{algorithmic}
\usepackage{booktabs}
\usepackage{caption}
\SIthousandsep{,}
\usepackage{makecell}
\usepackage[skip=0pt]{caption}
\usepackage{algorithm}
\usepackage{algorithmic}
\usepackage{amsmath}
\usepackage{subfigure}

%% file: definitions.tex
\AtBeginDocument{%
  \providecommand\BibTeX{{%
    \normalfont B\kern-0.5em{\scshape i\kern-0.25em b}\kern-0.8em\TeX}}}

\newcommand{\header}[1]{\vspace{1mm}\noindent\textbf{#1.}}

\DeclareMathOperator*{\argmax}{argmax}

\newcommand\blfootnote[1]{%
  \begingroup
  \renewcommand\thefootnote{*}\footnotetext{#1}%
  \addtocounter{footnote}{-1}%
  \endgroup
}

%% file: meta.tex
\copyrightyear{2022}
\acmYear{2022}
\setcopyright{rightsretained}
\acmDOI{}
\acmISBN{}

%% file: authors.tex
\author[Wu et al.]{
    Chen Wu\textsuperscript{\rm 1,2}, 
    Ruqing Zhang\textsuperscript{\rm 1,2},
    Jiafeng Guo\textsuperscript{\rm 1,2},
    Maarten de Rijke\textsuperscript{\rm 3},
    Yixing Fan\textsuperscript{\rm 1,2},
    and Xueqi Cheng\textsuperscript{\rm 1,2}
    \\
}

\affiliation{
  \institution{
  \textsuperscript{\rm 1}CAS Key Lab of Network Data Science and Technology, \\ Institute of Computing
Technology, Chinese Academy of Sciences; }
\country{}
}

\affiliation{
  \institution{\textsuperscript{\rm 2}University of Chinese Academy of Sciences}
  \country{}
} 

\affiliation{
  \institution{\textsuperscript{\rm 3}University of Amsterdam}
  \country{}
}

\email{{wuchen17z,zhangruqing,guojiafeng,fanyixing,cxq}@ict.ac.cn
}
\email{m.derijke@uva.nl}

\def\authors{Chen Wu, Ruqing Zhang, Jiafeng Guo, Maarten de Rijke, Yixing Fan, and Xueqi Cheng}